\documentclass[english,twocolumn,showpacs,preprintnumbers,amsmath,amssymb, prb, superscriptaddress]{revtex4}
\usepackage[T1]{fontenc}
\usepackage[latin9]{inputenc}
\usepackage{amstext}
\usepackage{graphicx}
\usepackage{amssymb}

\makeatletter
\@ifundefined{textcolor}{}
{%
 \definecolor{BLACK}{gray}{0}
 \definecolor{WHITE}{gray}{1}
 \definecolor{RED}{rgb}{1,0,0}
 \definecolor{GREEN}{rgb}{0,1,0}
 \definecolor{BLUE}{rgb}{0,0,1}
 \definecolor{CYAN}{cmyk}{1,0,0,0}
 \definecolor{MAGENTA}{cmyk}{0,1,0,0}
 \definecolor{YELLOW}{cmyk}{0,0,1,0}
 }

\usepackage{placeins}

\usepackage{dcolumn}

\usepackage{bm}

\makeatother

\usepackage{babel}

\begin{document}

\title{Heat conductivity of the spin-Peierls compounds TiOCl and TiOBr}

\author{N. Hlubek}

\affiliation{IFW-Dresden, Institute for Solid State Research,
P.O. Box 270116, D-01171 Dresden, Germany}

\author{M. Sing}

\author{S. Glawion}

\author{R. Claessen}

\affiliation{Experimentelle Physik 4, Universit\"at W\"urzburg,
Germany}

\author{S. van Smaalen}

\affiliation{Laboratory of Crystallography, Universit\"at Bayreuth,
Germany}

\author{P.H.M. van Loosdrecht}

\affiliation{Zernike Institute for Advanced Materials, University
of Groningen, Netherlands}

\author{B. B\"uchner}

\author{C. Hess}

\affiliation{IFW-Dresden, Institute for Solid State Research,
P.O. Box 270116, D-01171 Dresden, Germany}

\date{\today}

\pacs{66.70.-f, 75.40.Gb, 75.10.Pq, 68.65.-k}
\begin{abstract}
\noindent We report experimental results on the heat conductivity
$\kappa$ of the $S=1/2$ spin chain compounds TiOBr and TiOCl for
temperatures $5\,\mathrm{K}<T<300\,\mathrm{K}$ and magnetic fields
up to 14~T. Surprisingly, we find no evidence of a significant magnetic
contribution to $\kappa$, which is in stark contrast to recent results
on $S=1/2$ spin chain cuprates. Despite this unexpected result, the
thus predominantly phononic heat conductivity of these spin-Peierls
compounds exhibits a very unusual behavior. In particular, we observe
strong anomalies at the phase transitions $T_{c1}$ and $T_{c2}$.
Moreover, we find an overall but anisotropic suppression of $\kappa$
in the intermediate phase which extends even to temperatures higher
than $T_{c2}$. An external magnetic field causes a slight downshift
of the transition at $T_{c1}$ and enhances the suppression of $\kappa$
up to $T_{c2}$. We interprete our findings in terms of strong spin-phonon
coupling and phonon scattering arising from spin-driven lattice distortions. 
\end{abstract}
\maketitle

\section{Introduction}

Understanding low-dimensional quantum spin-1/2 systems is one of the
challenges of contemporary condensed matter physics. In particular,
transition metal oxides provide a rich playground for studying novel
phenomena, arising from the interplay between lattice, orbital, spin,
and charge degrees of freedom. The recent discovery of a substantial
magnetic heat conductivity $\kappa_{\mathrm{mag}}$ in 1D quantum
spin systems \citep{Sologubenko2000a,Sologubenko2001,Hess2001,Hess02,Hess04a,Hess05,Ribeiro05,Hess06,Hess2007,Hess2007b,Hlubek2010}
together with the theoretical prediction of ballistic transport in
1D $S=1/2$ Heisenberg chains \citep{Zotos1997,Kluemper2002,Heidrich2003}
has caused intense experimental and theoretical research on the behavior
of these systems. The best experimental realizations of $S=1/2$ systems
showing magnetic heat transport are up to now found among copper-oxides
(cuprates) such as the spin chains SrCuO$_{2}$ and Sr$_{2}$CuO$_{3}$
\citep{Sologubenko2001,Hlubek2010} and the spin ladder compounds
(Ca,La,Sr)$_{14}$Cu$_{24}$O$_{41}$.~\citep{Sologubenko2000a,Hess2001,Hess04a,Hess06}
Characteristic for the cuprate chain systems is a Cu3$d^{9}$ configuration
which gives rise to $S=1/2$ and a large exchange coupling $J/k_{B}\approx2000$~K
along the chain/ladder direction. As a consequence of this quasi 1D
magnetic structure, these systems exhibit a strongly anisotropic thermal
transport behavior. Perpendicular to the low-dimensional spin structure
a typical phononic heat conductivity $\kappa_{\mathrm{ph}}$ is found.
However, parallel to the low-dimensional spin structure the heat conductivity
is strongly enhanced even up to room temperature since a large $\kappa_{\mathrm{mag}}$
adds to $\kappa_{\mathrm{ph}}$. These in many aspects excellent realizations
of $S=1/2$ Heisenberg chains do not undergo a spin-Peierls transition,
i.e. a transition to a spin-dimerized ground state at the expense
of a lattice distortion that normally should arise from the spin-phonon
coupling of a spin chain and the phonon system in which it is embedded.
Surprisingly, only one Cu based spin system, CuGeO$_{3}$, is known
to exhibit a spin-Peierls transition \citep{Hase1993}. The exchange
energy of this compound is $J/k_{B}\approx160$~K \citep{Fabricius1998}
and the transition to the non-magnetic ground state is at $T_{c}\approx14$~K.
The heat conductivity of CuGeO$_{3}$ has been studied by several
groups \citep{Ando1998,Vasilev1997,Hofmann2002} with controversial
results. 1D magnetic heat conductivity has been suggested to give
rise to a significantly enhanced heat conductivity at $T<T_{c}$ \citep{Ando1998}.
However, the observed low-temperature peak has been shown to be present
both in the heat conductivity parallel and perpendicular to the chain
and thus can be rationalized in terms of phononic transport alone.~\citep{Hofmann2002}

\begin{figure*}[t]
\includegraphics[clip,width=8.8cm]{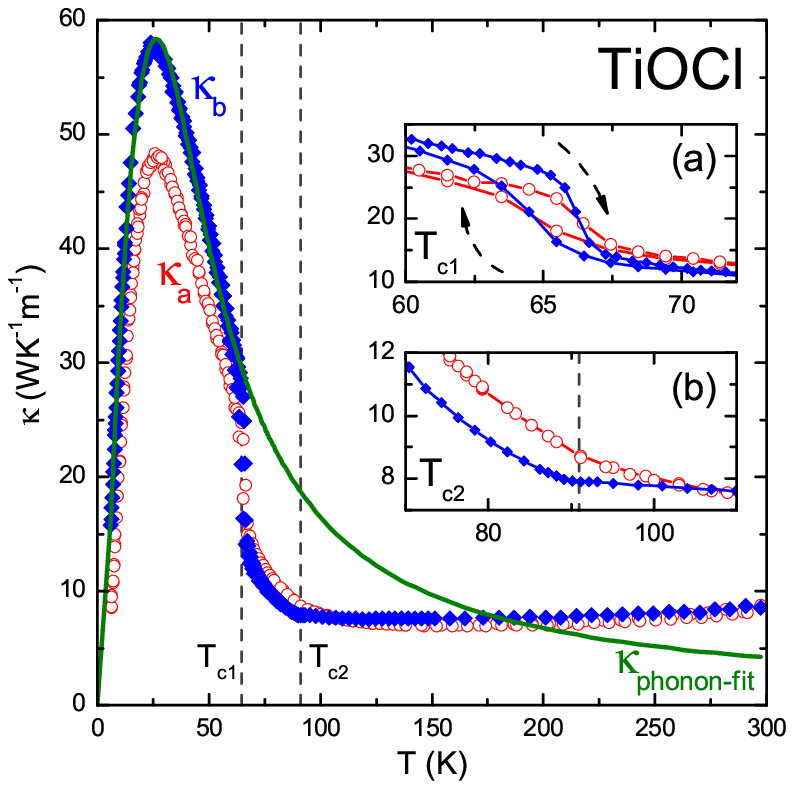}\hfill{}\includegraphics[clip,width=8.8cm]{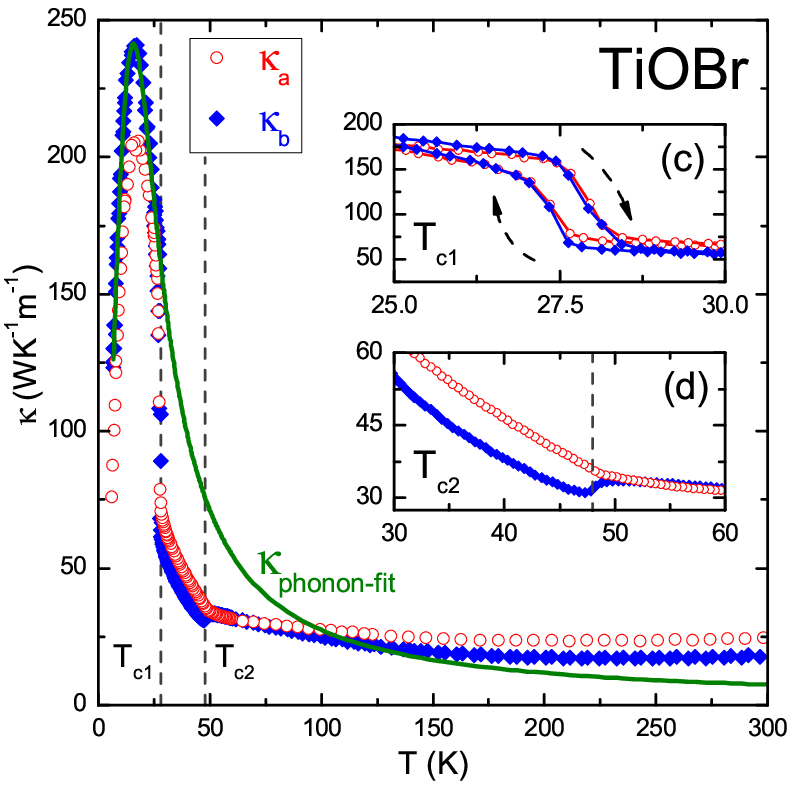}

\caption{(color online). Thermal conductivities $\kappa_{a}\left(\circ\right)$
and $\kappa_{b}\left(\blacklozenge\right)$ of TiOCl and TiOBr as
a function of $T$. The insets depict the behavior of the thermal
conductivity around the respective phase transitions. Insets (a) and
(c) show the hysteretic behavior around the phase transition at $T_{c1}$
which is characteristic for a first-order transition. The arrows mark
the corresponding curves for cooling and heating. The lower insets
(b) and (d) show $\kappa$ around the transition at $T_{c2}$ also
measured for cooling and heating. It is without hysteresis and therefore
the transition is of second order. \label{fig:Thermal-cond-of-TiOX}}

\end{figure*}

Also spin $S=1/2$ systems, but based on early transition metal ions
with electronic configuration 3$d^{1}$, the titanium oxyhalides TiOX,
with X=Br or Cl shifted recently into focus. These compounds are considered
as good realizations of $S=1/2$ spin chains which are formed by direct
overlap of Ti $t_{2g}$ orbitals along the crystallographic $b$ direction
\citep{Seidel2003,Kataev2003,Hoinkis2005} with rather high magnetic
exchange coupling {\small $J\left(\mathrm{Cl}\right)\approx676$~K~\citep{Seidel2003,Kataev2003}
and $J\left(\mathrm{Br}\right)\approx375$~K\citep{Rueckamp2005,Sasaki2005,Kato2005}}.
The compounds undergo two phase transitions $T_{c1}$, $T_{c2}$~\citep{Hemberger2005,Rueckamp2005}
where the lower one, at $T_{c1}$, leads to a non-magnetic dimerized
state~\citep{Seidel2003} which is accompanied by a doubling of the
unit cell.~\citep{Shaz2005,Sasaki2005} These features thus render
the Ti oxyhalides the second (besides CuGeO$_{3}$) type of inorganic
compounds which undergoes a spin-Peierls transition. However, as compared
to CuGeO$_{3}$ the dimerized state occurs at much higher temperatures,
viz. $T_{c1,\mathrm{Cl}}=67$~K for TiOCl and $T_{c1,\mathrm{Br}}=28$~K
for TiOBr. However, several experimental results are inconsistent
with a canonical spin-Peierls scenario. There are two successive phase
transitions and the transition to the non-magnetic state at $T_{c1}$
is of first \citep{Hemberger2005,Shaz2005,Sasaki2005} and not of
second order as in CuGeO$_{3}$. Interestingly, in the intermediate
regime between $T_{c1}$ and $T_{c2}$ an incommensurate superstructure
is found.~\citep{Abel2007,Smaalen2005,Clancy2007,Clancy2010} Above
$T_{c2}$ ($T_{c2,\mathrm{Cl}}=91$~K for TiOCl and $T_{c2,\mathrm{Br}}=48$~K
for TiOBr) the system is in a pseudo spin-gap regime up to a characteristic
temperature $T^{*}$ which for TiOCl extends up to $T^{*}\approx135$~K
with a large singlet-triplet energy gap of $E_{g}=430$~K.~\citep{Imai2003,Saha2007,Baker2007,Clancy2007}
First explanations of the intermediate phase proposed orbital fluctuations
but this has been ruled out by optical measurements in combination
with cluster calculations that showed, that the crystal field splitting
is large enough to quench the orbital degree of freedom.~\citep{Rueckamp2005}
Recent explanations focus on the interplay between intra- and interchain
frustrations and a related dimensionality crossover.~\citep{Abel2007,Fausti2007,Clancy2007,Macovez2007,Mastrogiuseppe2009,Zhang2008} 

The relatively high magnetic exchange constants of the titanium oxyhalides
render them good non-cuprate candidates for exhibiting a sizeable
magnetic heat conductivity arising from the 1D $S=1/2$ spin chains.
In this paper, we experimentally investigate the thermal conductivity
$\kappa$ of TiOCl and TiOBr with a special focus on potentially arising
magnetic contributions to $\kappa$. Surprisingly, no indication for
magnetic heat transport is observed and we find instead that $\kappa$
is dominated by phononic heat conduction. However, strong anomalies
occur at the phase transitions $T_{c1}$ and $T_{c2}$ and we find
an overall suppression of the phononic $\kappa$ which is anisotropic
in the incommensurate phase and which extends to temperatures higher
than $T_{c2}$. For TiOBr the application of an external magnetic
field of 14~T slightly shifts $T_{c1}$ towards lower temperature
and causes a weak further suppression of $\kappa$ in the intermediate
regime.

\section{Experiment}

Single crystals of TiOCl and TiOBr were synthesized by a chemical
vapor transport technique leading to small plate-like crystals.~\citep{Schaefer1958}
The crystallinity was checked by x-ray diffraction. Typical crystal
dimensions are a few $\mbox{m}\mbox{m}^{2}$ in the \textit{ab}-plane
but only around 20~$\mu\mbox{m}$ along the \textit{c} axis. Rectangular
samples with typical dimensions of $\left(2\times1\times0.02\right)\,\mbox{mm}^{3}$
with the longest side being parallel to the \textit{a} and \textit{b}
axis, respectively, were cut from the crystal plates. Measurements
of the thermal conductivity as a function of temperature $T$ in the
range of 7\textendash{}300 K were performed with a standard four probe
technique \citep{Hess2003b}. Because of the small thickness of the
crystals the usual uncertainty of 10\% for $\kappa$ due to the error
in the determination of the crystal geometry is exceeded by some extent.
Furthermore, the small thickness along the\textit{ c} axis also prevented
to measure $\kappa$ along this direction. In order to compare the
anisotropy of $\kappa$ along the $a$ and $b$ directions the individual
samples were cut from the same crystal plate thus keeping the relative
error between the two directions small. The mounting of TiOBr into
the heat conductivity probe was performed under Argon atmosphere in
order to minimize degradation of the sample.

\section{Results }

Figure~\ref{fig:Thermal-cond-of-TiOX} shows the temperature dependence
of the thermal conductivities along the \textit{a} and \textit{b}
axes ($\kappa_{a}$ and $\kappa_{b}$) of TiOCl and TiOBr in zero
magnetic field. We focus first on the results for TiOCl which are
shown in the left panel of Fig.~\ref{fig:Thermal-cond-of-TiOX}.
A first glance at the data already suggests that the temperature dependence
of $\kappa$ is governed by the two phase transitions at $T_{c1}$
and $T_{c2}$ which divide the data into three regimes. At low temperature
the heat conductivity parallel to the chains, $\kappa_{b}$, exhibits
a strong peak at $\sim25$~K with a maximum value $\kappa_{b}\thickapprox58\,\mathrm{Wm^{-1}K^{-1}}$
which is a typical feature of a phononic heat conductivity $\kappa_{\mathrm{ph}}$
at low temperature. It arises from two competing effects: At very
low temperature the mean free path of phonons is determined by the
crystal boundaries and defects and therefore is practically $T$-independent.
Hence, $\kappa_{\mathrm{ph}}$ increases due to the increasing number
of phonons. At higher temperature the mean free path is $T$-dependent
as the number of umklapp processes rises exponentially. This overcompensates
the effect of a rising phonon population and thus $\kappa_{\mathrm{ph}}$
decreases with further rising $T$, i.e., $\kappa_{\mathrm{ph}}$
usually shows a maximum. Interestingly, $\kappa_{b}$ deviates from
this conventional behavior at $T_{c1}$, where a sharp drop occurs
to about 60\% of the value of $\kappa_{b}$ at just below the transition.
In the intermediate phase $\kappa_{b}$ continuously decreases further
with rising $T$ just until $T_{c2}$ is reached. Upon rising $T$
through $T_{c2}$ we find that $\kappa_{b}$ changes slope and exhibits
a weak increase in the entire high temperature phase, i.e., at $T>T_{c2}$,
up to room temperature.

A very similar temperature dependence is observed in the heat conductivity
perpendicular to the chains, $\kappa_{a}$. In this case, the peak
at $T\approx25$~K is somewhat smaller ($\kappa_{a,\mathrm{max}}\approx48\,\mathrm{Wm^{-1}K^{-1}}$)
than that in $\kappa_{b}$. A similarly sharp drop as in the latter
occurs at $T_{c1}$. However, the actual drop at the transition is
relatively weaker as in the other direction. Interestingly, despite
a similar further decrease of $\kappa_{a}$ when rising $T$ towards
$T_{c2}$ as in $\kappa_{b}$, we find that $\kappa_{a}$ remains
always somewhat \textit{larger} in this intermediate regime. The slope
of $\kappa_{a}$ changes at $T_{c2}$, but remains negative up to
$T\approx150$~K, in contrast to the findings for $\kappa_{b}$ (cf.
Fig.~\ref{fig:Thermal-cond-of-TiOX}(b)).

Before discussing these pecularities in detail, we briefly summarize
the results for TiOBr which are shown in the right panel of Fig.~\ref{fig:Thermal-cond-of-TiOX}.
The general $T$-dependence of $\kappa$ has large similarities with
that of TiOCl, including the observed anomalies. There are, however,
slight differences which are worth to be pointed out: First, the phononic
peak of both $\kappa_{a}$ and $\kappa_{b}$ of TiOBr is by a factor
of about 4 larger than that in TiOCl and is located at somewhat lower
temperature ($\sim17$~K). Both features point to a lower defect
density in the case of TiOBr. This is corroborated by room temperature
x-ray diffraction which showed much sharper spots for TiOBr. Second,
at $T<T_{c2}$ the anisotropy between $\kappa_{a}$ and $\kappa_{b}$
is similar to that of TiOCl. More specifically, at $T<T_{c1}$ we
find $\kappa_{a}<\kappa_{b}$, and $\kappa_{a}>\kappa_{b}$ at $T_{c1}<T<T_{c2}$,
i.e. the drop at $T_{c1}$ and the reduction of $\kappa$ are relatively
stronger in $\kappa_{b}$ than that in $\kappa_{a}$. Interestingly,
the anomaly in $\kappa_{b}$ at $T_{c2}$ is much stronger than that
in TiOCl since a clear dip is observable at the transition (cf. Fig.~\ref{fig:Thermal-cond-of-TiOX}(d)).
Moreover, in contrast to TiOCl we observe that both $\kappa_{a}$
and $\kappa_{b}$ decrease with rising temperature at $T>T_{c2}$
up to room temperature where $\kappa_{a}$ remains slightly larger
than $\kappa_{b}$.

\section{Discussion}

The overall very weak anisotropy of the $\kappa$ data suggests without
further analysis the unexpected conclusion that magnetic heat transport
in the spin chains of this material is negligible in both TiOCl and
TiOBr. Otherwise a significant enhacement of $\kappa_{b}$ with respect
to $\kappa_{a}$ should occur since heat transport by magnetic excitations
is only expected along the 1D spin chain, i.e. parallel to $b$. One
might speculate that the weak anisotropy that is present in the low
temperature regime $T<T_{c1}$ is the indication of a weak magnetic
contribution along $b$ which could give rise to the observed $\kappa_{b}>\kappa_{a}$.
However, the observed anisotropy by a factor $\sim1.2$ matches that
of other phononic heat conductors \citealp{Hess2003b,Hess1999} and
can conventionally be explained by differences in the phonon velocity. 

At higher temperatures $\left(T>T_{c1}\right)$ magnetic contributions
appear even more unlikely, since in all cases $\kappa_{b}\lesssim\kappa_{a}$.
However, in this regime a small magnetic contribution to $\kappa_{b}$
might still be present if the expected anisotropy was masked by differences
in the phononic transport along the two crystallographic directions.
Concentrating only on the thermal conductivity $\kappa_{b}$ we estimate
the thus maximum possible $\kappa_{\mathrm{mag}}$ by performing a
phononic fit based on the so-called Callaway model \citep{Callaway1959}
to the low temperature peak and extrapolate this fit towards room
temperature. The fit is depicted by the solid line in Fig.~\ref{fig:Thermal-cond-of-TiOX}
and yields a very good agreement up to $T_{c1}$ but deviates strongly
from the data at higher temperatures. In particular, at high temperatures
($T\gtrsim180$~K) the fit is clearly lower than the data. We use
the difference between the fit $\kappa_{\mathrm{ph,}\mathrm{Fit}}$
and the data at room temperature to obtain an upper estimate for the
possible magnetic contributions $\kappa_{\mathrm{mag}}=\kappa_{b}-\kappa_{\mathrm{ph,}\mathrm{Fit}}$.
In order to analyse the thermal transport we estimate the magnetic
mean free path $l_{\mathrm{mag}}$ using an approximation of $\kappa_{\mathrm{mag}}$
of a $S=1/2$ Heisenberg chain \citealp{Hess2007b,Sologubenko2001}
\begin{equation}
\kappa_{\mathrm{mag}}=\frac{2n_{s}{k_{B}}^{2}}{\pi\hbar}l_{\mathrm{mag}}T\int_{0}^{\frac{J\pi}{2k_{B}T}}x^{2}\frac{\exp(x)}{(\exp(x)+1)^{2}}dx,\end{equation}
where $n_{s}$ is a geometrical factor that counts the number of chains
per unit area. For both compounds this yields a negligibly small mean
free path of only 2-3 lattice constants. %
\footnote{A different approach to estimate $\kappa_{\mathrm{mag}}$ that follows
an analysis described in Ref.~\onlinecite{Sologubenko2003b} using
the thermal Drude weight \citep{Kluemper2002} yields the same result. %
} Considering the fact that the Callaway model usually underestimates
$\kappa_{\mathrm{ph}}$ at room temperature \citealp{Berman65,Asen-Palmer97}
and that $\kappa_{b}\lesssim\kappa_{a}$ at higher temperature any
realistic value for the mean free path should be even smaller which
essentially rules out magnetic transport in the Ti oxyhalides.

There are not many scenarios which straightforwardly explain this
unexpected result. The absence of magnetic heat conduction in magnetic
materials has been discussed by Sanders and Walton in terms of a very
large magnon-phonon relaxation time \citealp{Sanders77}. It is obvious
that this situation cannot be realized in Ti-oxyhalides since a significant
spin-phonon coupling must be present in these compounds to allow for
a spin-Peierls transition at considerably high temperatures. In fact,
it is therefore more reasonable to explain the absence of magnetic
heat conduction by a particularly strong spin-phonon coupling which
gives rise to strong scattering of spin excitations and thus prevents
the magnetic heat conduction. One might speculate that even more exotic
excitations such as orbital fluctuations are relevant for suppressing
$\kappa_{\mathrm{mag}}$. We point out, however, that orbital excitations
have been shown to be unimportant for the low-energy physics in these
compounds.~\citep{Rueckamp2005,Hoinkis2005}

\begin{figure}[t]
\includegraphics[clip,width=8.8cm]{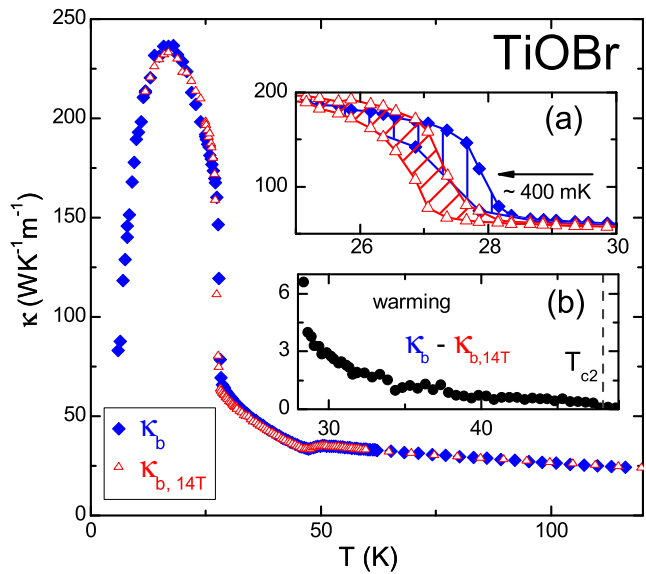}

\caption{(color online). Thermal conductivity $\kappa_{b}$ as a function of
$T$ in TiOBr with ($\kappa_{b,14T}\left(\vartriangle\right)$) and
without ($\kappa_{b}\left(\blacklozenge\right)$) an applied magnetic
field of 14T along the chain direction. Inset (a) illustrates the
shift of $T_{c1}$ towards lower temperatures in the presence of a
magnetic field. In inset (b) $\text{\ensuremath{\Delta\kappa}}=\kappa_{b}-\kappa_{b,14\mathrm{T}}$
shows the decreasing influence of the magnetic field on $\kappa_{b}$
in the intermediate regime. The curves used in the substraction are
from the measurements that approach the phase transitions from low
temperatures. \label{fig:Thermal-conductivity-in-field}}

\end{figure}

The negligible magnetic heat conduction in the Ti-oxyhalides implies
that the unusual temperature dependence and also the slight anisotropy
should be rationalized in terms of pure phonon heat conduction, which
has been proven to be a sensitive probe to pecularities of the lattice
such as superstructures and disorder.~\citealp{Hess2003b,Hess1999,Berggold2008,Cassel99a}
The considerable jump in $\kappa$ at $T_{c1}$ clearly indicates
that the phonon heat conduction in the intermediate phase is strongly
suppressed with respect to that of the commensurate dimerized phase
at $T<T_{c1}$ where ordinary phonon heat conduction is observed.
This reflects the abrupt transition towards a lattice with strongly
disturbed periodicity and anharmonicity which causes enhanced phonon
scattering and is entirely consistent with the incommensurate lattice
distortion in this regime.~\citep{Imai2003,Smaalen2005,Schoenleber2006}
We have investigated the nature of this phase transition at $T_{c1}$
and find for both compounds a clear hysteretic behavior which confirms
the transition being of first order (see Fig. ~\ref{fig:Thermal-cond-of-TiOX}(a)
and \ref{fig:Thermal-cond-of-TiOX}(c)). Such first-order character
has already been reported from magnetic susceptibility, specific heat,
thermal expansion and x-ray data of the superstructure satellites.~\citep{Hoinkis2005,Hemberger2005,Rueckamp2005,Shaz2005}
Since the magnetic exchange is smallest in TiOBr we have searched
for possible effects of a magnetic field on $\kappa_{b}$. As is depicted
in Fig.~\ref{fig:Thermal-conductivity-in-field} a magnetic field
of $B=14$~T along the \textit{b} direction has only little influence
on the thermal conductivity $\kappa_{b,14\mathrm{T}}$. However, we
detect a slight downshift of the phase transition at $T_{c1}$ by
$\sim400\,\mathrm{mK}$ which is consistent with a downshift of $\sim130\,\mathrm{mK}$
that has been reported from x-ray diffraction at $B=10$~T for TiOCl.~\citep{Krimmel2006}
Moreover, starting at $T_{c1}$, $\kappa_{b,14\mathrm{T}}$ is slightly
smaller compared to the measurement without field, but gradually approaches
it for increasing temperature. In Fig.~\ref{fig:Thermal-conductivity-in-field}(b)
the difference $\Delta\kappa=\kappa_{b}-\kappa_{b,14\mathrm{T}}$
between both curves is shown, illustrating the decreasing influence
of the magnetic field, until it vanishes at $T_{c2}$. This suggests
that the spin-induced incommensurate lattice distortion in this intermediate
phase is further enhanced by an external magnetic field.

The thermal conductivity across the phase transitions at $T_{c2}$
shown in more detail in Fig.~\ref{fig:Thermal-cond-of-TiOX}(b) and
\ref{fig:Thermal-cond-of-TiOX}(d), does not exhibit a hysteretic
behavior which is indicative of a second-order transition. The overall
impact of this transition on $\kappa$ is much smaller than that at
$T_{c1}$. Interestingly, in the high-temperature phase above $T_{c2}$
the thermal conductivity appears still significantly suppressed with
respect to the low-temperature phase at $T<T_{c1}$. In Fig.~\ref{fig:Thermal-cond-of-TiOX}
this is clearly seen when comparing the data to the phononic fit which
remains much larger than $\kappa$ up to $T^{*}\sim100$~K and $T^{*}\sim180$~K
for TiOBr and TiOCl, respectively. Only at higher temperatures a more
typical behavior is observed with $\kappa_{\mathrm{ph,}\mathrm{Fit}}<\kappa$.
The apparent suppression of $\kappa$ in the regime $T_{c2}\lesssim T\lesssim T^{*}$
clearly indicates, that strong phonon scattering occurs despite the
absence of any static long range lattice distortions. A reasonable
origin of this enhanced scattering are precursors of the spin-Peierls
transition, either as short-range static lattice distortions or as
slowly fluctuating precursors (soft phonon type). This is consistent
with the pseudogap seen in magnetic resonance measurements \citep{Saha2007,Imai2003,Kataev2003}
and incommensurate structural fluctuations found by x-ray diffraction.~\citep{Schoenleber2006} 

In both compounds the suppression of $\kappa$ in the intermediate
phase is clearly anisotropic, since the drop of $\kappa_{b}$ at $T_{c1}$
is relatively stronger as compared to $\kappa_{a}$ and $\kappa_{b}$<$\kappa_{a}$
in the entire phase where $\kappa_{b}$ of TiOBr even shows a local
minimum at $T_{c2}$. Similar anisotropic scattering has previously
been observed, e.g., in stripe-ordering compounds which possess anisotropic
correlation lengths of the stripe order close to the transition.~\citep{Hess1999}
The stronger suppression of $\kappa_{b}$ than $\kappa_{a}$ in the
present case can be understood by looking at the modulation amplitudes
for TiOCl \citep{Schoenleber2006} and TiOBr in the incommensurate
phase \citep{Smaalen2005}. Those indicate that the shifts of the
atoms out of the periodic position of the structure at room temperature
are larger in the direction of the \textit{b} axis than those along
the \textit{a} axis. The resulting larger anharmonicity along $b$
is likely causing increased scattering and therefore the observed
lower thermal conductivity.

There is a slight difference in the thermal conductivity between both
compounds near room temperature where phonon scattering arising from
the spin-Peierls transition can be considered to be relatively weak.
For TiOBr the slope of $\kappa$ is negative while it is positive
for TiOCl. At the same time the absolute value of $\kappa$ is significantly
higher in TiOBr. This corroborates the previous conclusion that our
TiOBr crystals have a lower defect density than the TiOCl ones because
the observed temperature dependence of $\kappa$ for TiOBr is much
closer to the expected $\propto T^{-1}$ decrease of a clean phonon
heat conductor.~\citealp{Berman} On the other hand, the lower $\kappa$
of TiOCl with a weak positive slope is typical for more disordered
heat conductors, where also rather small contributions to $\kappa$,
such as heat transport by optical phonons\citep{Hess2004} become
relevant.

\section{Summary}

In conclusion, we have shown that the magnetic thermal conductivity
in the TiOX is negligible due to strong spin-phonon scattering. The
heat transport can thus be understood in terms of pure phononic conductivity.
At the phase transitions we find strong anomalies which are consistent
with the lattice distortions. Starting at low temperatures, the first
phase transition $T_{c1}$ towards the dimerized state can be shifted
to lower temperatures by an external magnetic field. Additionally,
this leads to a slight suppression of the thermal conductivity throughout
the intermediate regime and gradually gets smaller when approaching
$T_{c2}$. Comparing the measurements along the different crystallographic
directions in this regime, the stronger suppression along $\kappa_{b}$
for both compounds is consistent with a higher incommensurability
of the lattice in this direction. Finally, by a comparison of the
extrapolated thermal conductivity from a phononic model to the measurement
at higher temperatures it was argued that the thermal conductivity
is still supressed up to a temperature $T^{*}$ which is either a
sign of short-range lattice distortions or phonon softening. 

\begin{acknowledgments}
We thank Daniel Khomskii and Daniele Fausti for valuable discussions.
This work was supported by the Deutsche Forschungsgemeinschaft through
grant HE3439/7, SM55/15 and CL124/6, through the Forschergruppe FOR912
(grant HE3439/8) and by the European Commission through the NOVMAG
project (FP6-032980). 
\end{acknowledgments}

\end{document}